\documentstyle[epsf]{l-aa}

\begin{document}

\thesaurus{03(03.04.1; 
              07.09.1; 
              07.19.1: 
              09.13.1; 
              20.01.2) 
                      }
\title 
{Extragalactic Large-Scale Structures behind the Southern Milky Way.
-- II. Redshifts Obtained at the SAAO in the Crux region
}

\author{A.P.~Fairall\inst{1} \and P.A.~Woudt\inst{1} 
\and R.C.~Kraan-Korteweg\inst{2} }

\offprints {A.P. Fairall}

\institute{
Department of Astronomy, University of Cape Town,
Rondebosch, 7700 South Africa
\and
Observatoire de Paris, DAEC, Unit\'e associ\'ee au CNRS, D0173, et \`a
l'Universit\'e Paris 7, 92195 Meudon Cedex, France}

\date{Received date; Accepted date}
\maketitle
\markboth{A.P.~Fairall et~al.: Large-Scale Structures behind the
Southern Milky Way. - II.}{Fairall et al.}

\def\sun{\hbox{$\odot$}}
\def\earth{\hbox{$\oplus$}}
\def\la{\mathrel{\hbox{\rlap{\hbox{\lower4pt\hbox{$\sim$}}}\hbox{$<$}}}}
\def\ga{\mathrel{\hbox{\rlap{\hbox{\lower4pt\hbox{$\sim$}}}\hbox{$>$}}}}
\def\sq{\hbox{\rlap{$\sqcap$}$\sqcup$}}
\def\arcmin{\hbox{$^\prime$}}
\def\arcsec{\hbox{$^{\prime\prime}$}}
\def\fd{\hbox{$.\!\!^{d}$}}
\def\fh{\hbox{$.\!\!^{h}$}}
\def\fm{\hbox{$.\!\!^{m}$}}
\def\fs{\hbox{$.\!\!^{s}$}}
\def\fdg{\hbox{$.\!\!^\circ$}}
\def\farcm{\hbox{$.\mkern-4mu^\prime$}}
\def\farcs{\hbox{$.\!\!^{\prime\prime}$}}
\def\fp{\hbox{$.\!\!^{\scriptscriptstyle\rm p}$}}
\def\micron{\hbox{$\mu$m}}

\newcommand{\etal}{{\it et al.}\,}      
\newcommand{\eg}{{\it e.g.},\ }         
\newcommand{\ie}{{\it i.e.},\ }         
\newcommand{\cf}{{\it cf.},\ }          
\newcommand{\CDOT}{$\cdot$}             
%
%
\newcommand{\Xray}{\hbox{X-ray}}
\newcommand{\Xrays}{\hbox{X-rays}}
%
%
\def\deg{{^\circ}}
\newcommand{\Deg}{$^\circ$}
\newcommand{\MAG}{$^m\llap{.\thinspace}$}
\newcommand{\Mag}{^m\llap{.\thinspace}}
\newcommand{\rsun}{\,{\rm R}_\odot}
\newcommand{\RSUN}{${\rm R}_\odot$}
\newcommand{\msun}{\,{\rm M}_\odot}
\newcommand{\MSUN}{${\rm M}_\odot$}
\newcommand{\lsun}{\,{\rm L}_\odot}
\newcommand{\LSUN}{${\rm L}_\odot$}
\newcommand{\vsun}{\,v_\odot}
\newcommand{\VSUN}{$v_\odot$}
\newcommand{\kms}{{\,km\ s$^{-1}$\,}}
\newcommand{\KMS}{{$km\ s^{-1}$}}
\newcommand{\HA}{\mbox{H$\alpha$}}
\newcommand{\LA}{\mbox{Lyman-$\alpha$}}
\newcommand{\NII}{\mbox{\normalsize [N\thinspace\footnotesize II\normalsize ]}}
\newcommand{\NIII}{\mbox{\normalsize [N\thinspace\footnotesize III\normalsize ]}}
\newcommand{\OI}{\mbox{\normalsize [O\thinspace\footnotesize I\normalsize ]}}
\newcommand{\OII}{\mbox{\normalsize [O\thinspace\footnotesize II\normalsize ]}}
\newcommand{\OIII}{\mbox{\normalsize [O\thinspace\footnotesize III\normalsize ]}}
\newcommand{\SII}{\mbox{\normalsize [S\thinspace\footnotesize II\normalsize ]}}
\newcommand{\HI}{\mbox{\normalsize H\thinspace\footnotesize I}}
\newcommand{\HII}{\mbox{\normalsize H\thinspace\footnotesize II}}
\newcommand{\MHI}{$M_{HI}$}
\newcommand{\MT}{$M_T$}
\newcommand{\BT}{$B_T^{0,i}$}
\newcommand{\MBT}{$M_{B_T}^{0,i}$}
\newcommand{\SHI}{$\sigma_{HI}$}
\newcommand{\MHILB}{$M_{HI}/L_B$}
\newcommand{\MHIMT}{$M_{HI}/M_T$}
\newcommand{\MTLB}{$M_T/L_B$}
\newcommand{\DVT}{$\Delta v_{20}$}
\newcommand{\DVF}{$\Delta v_{50}$}
\newcommand{\DVI}{$\Delta v_{0,i}$}
\newcommand{\LDV}{$\log\Delta v_{0,i}$}

\begin{abstract}
In our systematic optical galaxy search behind the southern Milky Way,
3760 (mostly unknown) galaxies with diameters D$\ga0\farcm2$ were 
identified in the Crux region ($287\deg \la \ell \la
318\deg, |b| \la 10\deg$, Woudt \& Kraan-Korteweg 1997). 
Prior to this investigation, only 65 of these galaxies had known 
redshifts. In order to map the galaxy 
distribution in redshift space we obtained spectra for 226
bright (B$_{\rm{J}} \la 18\mag0$) objects with the 1.9m 
telescope of the South African Astronomical Observatory (SAAO). 

Redshifts could be determined for 209 objects, of which
173 have good signal-to-noise ratios. 
Of the 36 tentative redshifts, four are confirmed 
through independent values in the literature. The redshifts of three objects 
indicate them to be galactic of origin. One of these confirms a 
suspected Planetary Nebula. For 17 of the galaxies, no redshift could 
be determined due to poor signal-to-noise ratios. 

In addition, 26 redshifts have have been measured in the Hydra-Antlia 
region investigated earlier (Kraan-Korteweg, Fairall \& Balkowski 1995), of
which one is a tentative estimate.

Two main structures crossing the Galactic Plane in the Crux region have now 
become clear. A narrow, nearby filament from ($\ell, b$) = ($340\deg, -25\deg$) 
to the Centaurus cluster can be traced. This filament runs almost parallel to 
the extension of the Hydra--Antlia clusters found earlier and is part of what we
have earlier termed the ``Centaurus Wall'' extending in redshift-space
between $0 \le v \le 6000$ \kms 
(Fairall \& Paverd 1995). The main outcome of this survey however, 
is the recognition of another massive extended structure between
$4000 \le v \le 8000$ {\kms}.
This broad structure, dubbed the Norma Supercluster (Woudt et al. 1997),
runs nearly parallel to the Galactic Plane from Vela to ACO 3627 (its centre) 
from where it continues to the Pavo cluster.
This massive structure is believed to be associated with the Great Attractor.

The survey has furthermore revealed a set of cellular structures, similar to those
seen in redshift space at higher galactic latitudes, but never before seen
so clearly behind the Milky Way. 

\end{abstract}


\section{Introduction}

This paper is the second in a series reporting redshifts for galaxies 
found at very low galactic latitudes, obtained for the purpose of mapping 
large-scale structures previously hidden by the southern Milky Way.  
The first paper (Kraan-Korteweg, Fairall \& Balkowski 1995 - hereafter 
Paper I) dealt with the Hydra--Antlia region; it also gave full details 
of the nature of the survey.  
The present paper covers the ``Crux'' region immediately neighbouring 
Hydra--Antlia; it does not repeat information given in Paper I, to 
which the reader is referred for an extensive description of the 
motivation, survey and observing procedures.
 
As before, galaxies in the general galactic latitude 
range $|b| \la 10\deg$ have been found by scanning film copies of 
the SRC IIIaJ survey under 50 times magnification, the details of which 
will be presented as a catalogue (Woudt \& Kraan-Korteweg 1997, hereafter WKK97).  
For the present work, some 500 square degrees of sky has been searched.  
It involves 21 fields limited at $|b| = 10\deg$, 
namely F62-67, F94-98, F130-134 and F171-175.

Figure 1 shows the distribution of the 3760 galaxies found in the
Crux region. Only $\sim 2.3$\% (= 87) of these galaxies had been 
catalogued before by Lauberts (1982). The entire area surveyed so far 
by us is indicated in Figure 1. The dashed line on the right demarcates
the Hydra--Antlia region, the middle section (solid line) is the Crux region
whereas the dashed line on the left shows the Great Attractor region.

\begin{figure*}
\hfil\epsfxsize=16.5cm \epsfbox{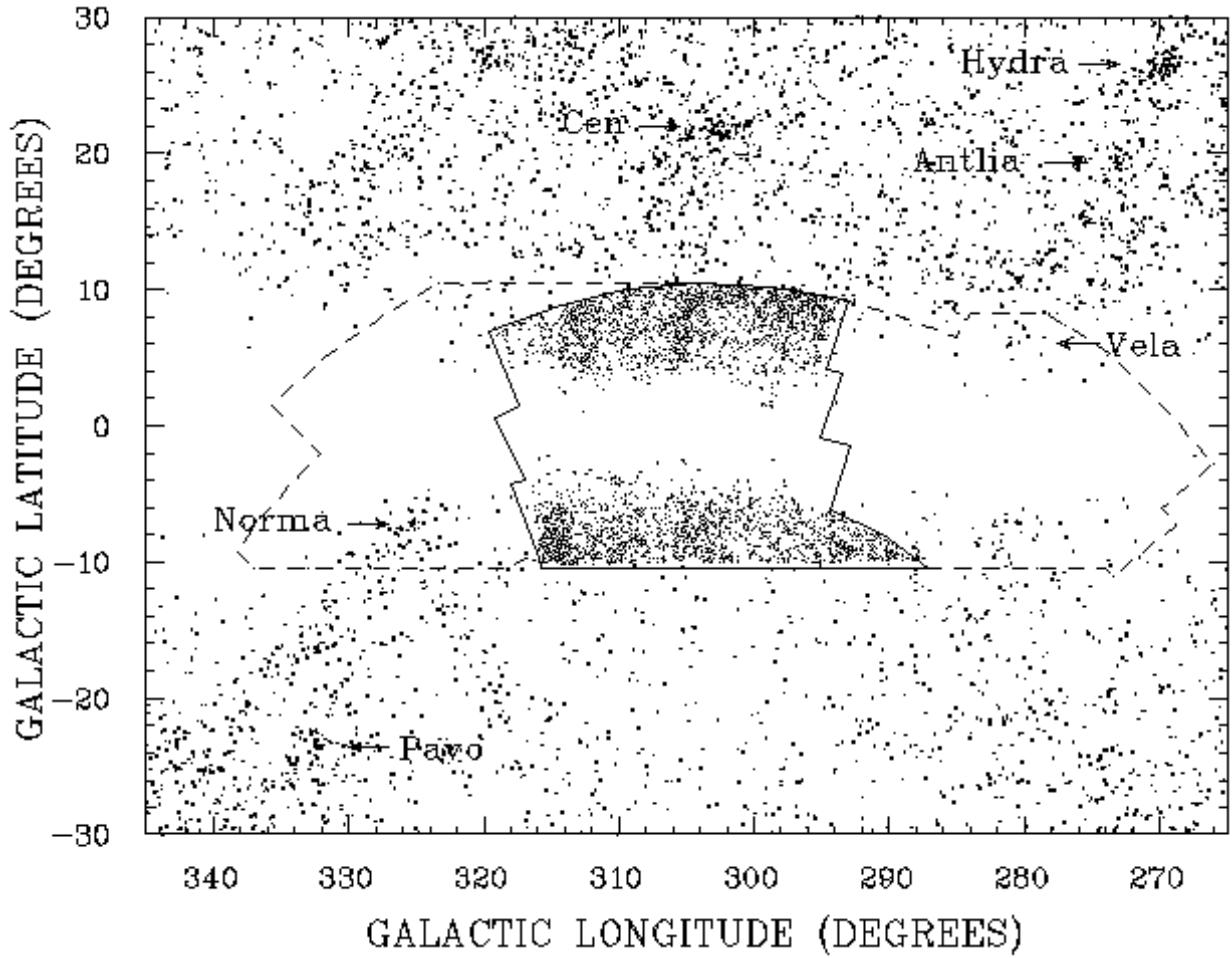}\hfil
\caption
{The distribution in Galactic coordinates of the galaxies in the Crux area.
The surveyed area is indicated and the newly identified galaxies with
$D \ge 0.2$ arcminutes entered as small dots. The larger dots in the 
surrounding area display the Lauberts galaxies ($D \ge 1.0\arcmin$). 
The Hydra, Antlia, Centaurus, Pavo and ACO 3627 (=Norma) clusters are 
labelled, as is the position of the Vela overdensity (reported in Paper I).
}
\label{f1}
\end{figure*}

The newly found galaxies in the Crux region significantly reduce 
the width of the ``Zone of Avoidance'' (ZOA) created by the Milky Way. 
Nevertheless there is a generally sharp cutoff along the northern 
boundary at $b = +3\deg$ and a more gradual but irregular extinction 
between $b = -3\deg$ and $-5\deg$ on the southern edge.  
The forthcoming blind HI-survey with the multi-beam system at Parkes should
unveil the remaining parts of the extragalactic sky between these boundaries.
 
The distribution of optically detected galaxies inside the Crux region is far 
from smooth. North of the Milky Way, there are three apparent concentrations 
at galactic longitudes $\ell = 296\deg, 305\deg$ and $313\deg$. The latter two 
suggest a continuation of the Centaurus Wall into the ZOA. South of the Milky Way, 
there is a concentration (at $\ell = 315\deg$) which is most likely a distant 
overdensity as the survey reveals a corresponding excess of smaller 
fainter galaxies.

The Crux region does include part of the ``Great Attractor'' (GA) (Lynden-Bell et al. 1987).
However, the main component of the GA lies within the boundaries of our 
neighbouring search area ($318\deg \la \ell \la 340\deg, |b| \la 10\deg$)
including the centre of the GA which Kolatt et al. (1995) predict 
at $\ell = 320\deg$, $b = 0\deg$. 
That region will be covered in a third paper in this series; it is dominated 
by ACO 3627 -- hereafter named the Norma cluster -- already recognised 
as a nearby massive cluster (Kraan-Korteweg et al. 1996 - hereafter KK96) 
while work on this survey was in progress.
 
\section {Observations}

As before, spectroscopic observations of the more widespread 
brighter (B$_{\rm{J}} \la 18\mag0$) galaxies, identified in our survey, 
have been carried out with the 1.9 m Radcliffe reflector and UNIT 
Spectrograph at the South African Astronomical Observatory during four
weeks in 1993 -- 1995, and are reported in Section 2 below.  
We have also continued 
to complement these observations by programmes using the MEFOS 
(multi-fibre) spectrograph system (Felenbok et al. 1997) 
on the 3.6m telescope at the 
European Southern Observatory (for fainter distant galaxies), 
and using the Parkes radio telescope (to cope with nearby galaxies 
of low optical brightness).  The outcome of these complementary 
programmes will be presented separately.

The observations at SAAO were carried out over 3500 -- 
7000{\AA} with a resolution of $\sim${3{\AA}} per pixel. 
Integration times typically range from 500 to 2500 seconds.
The procedures used for observations, and the reductions 
carried out thereafter at the University of Cape Town, are the same 
as described in Paper I.

\begin{table*}
\caption{Galaxies with redshifts obtained at S.A.A.O. in the Crux Region.}
\hfil\epsfxsize 17.5cm \epsfbox[30 6 610 749]{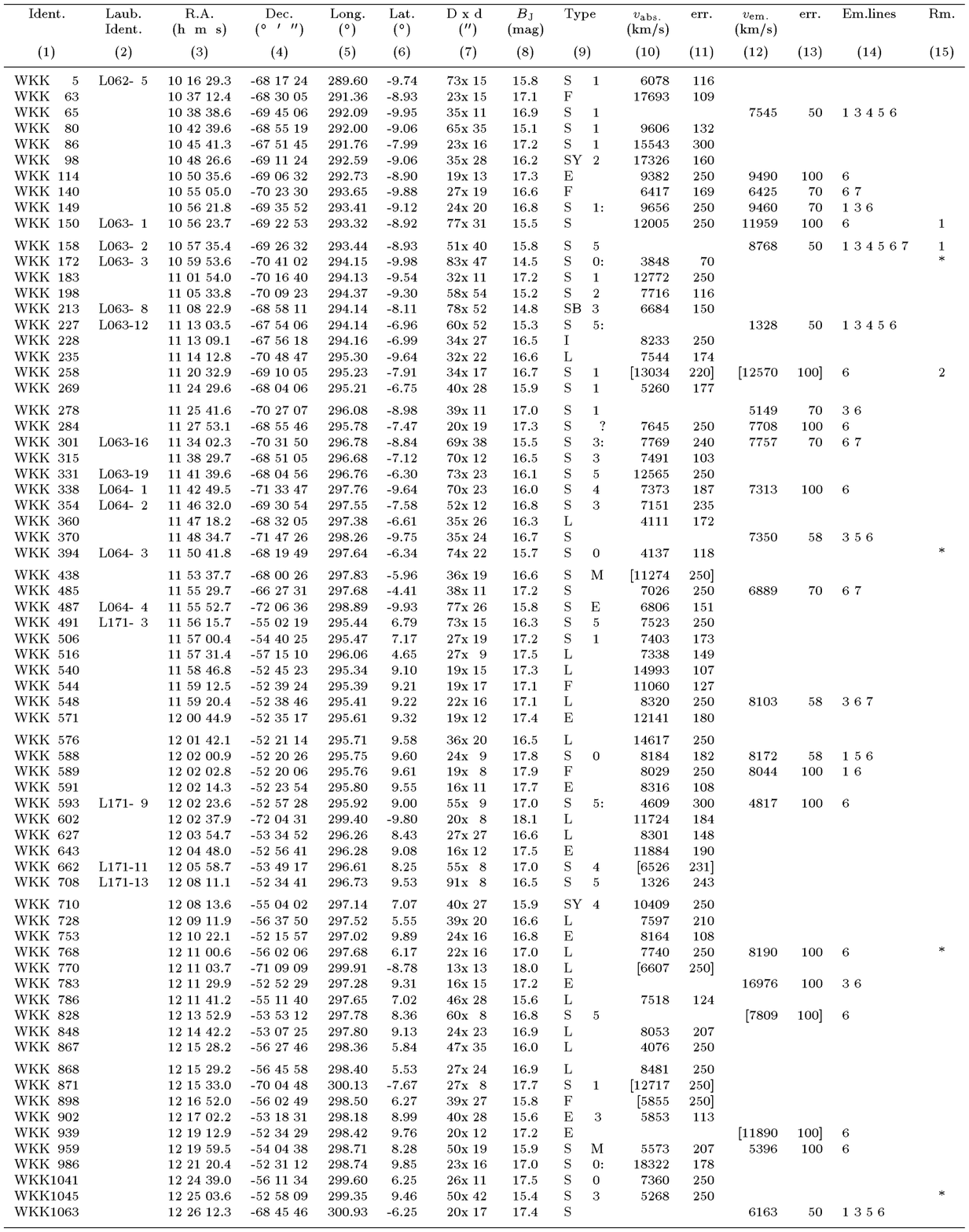}\hfil 
\label{t1}
\end{table*}
\addtocounter{table}{-1}
 
\begin{table*}
\caption{-- continued}
\hfil\epsfxsize 17.5cm \epsfbox[30 6 610 749]{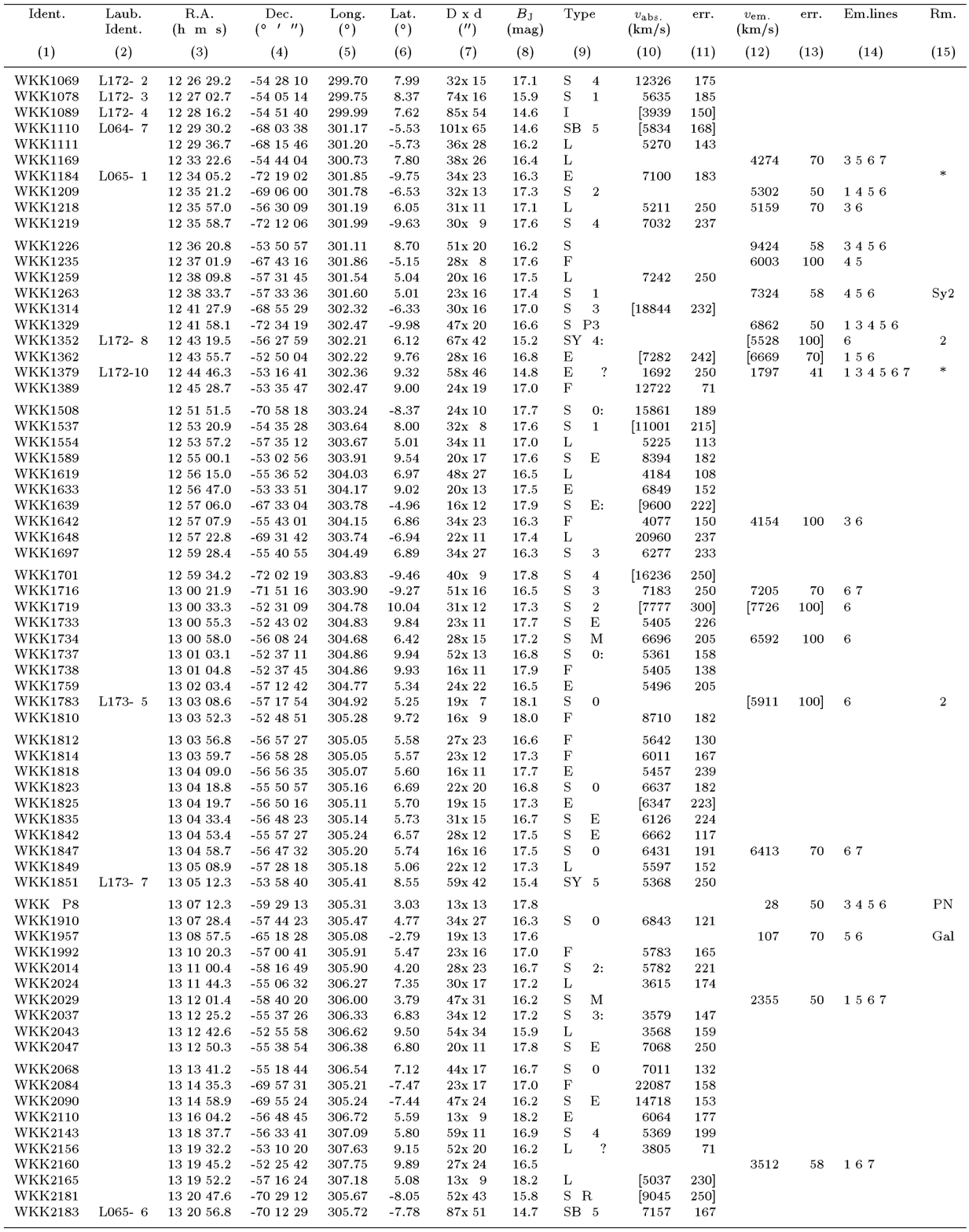}\hfil
\end{table*}
\addtocounter{table}{-1}
 
\begin{table*}
\caption{-- continued}
\hfil\epsfxsize 17.5cm \epsfbox[30 6 610 749]{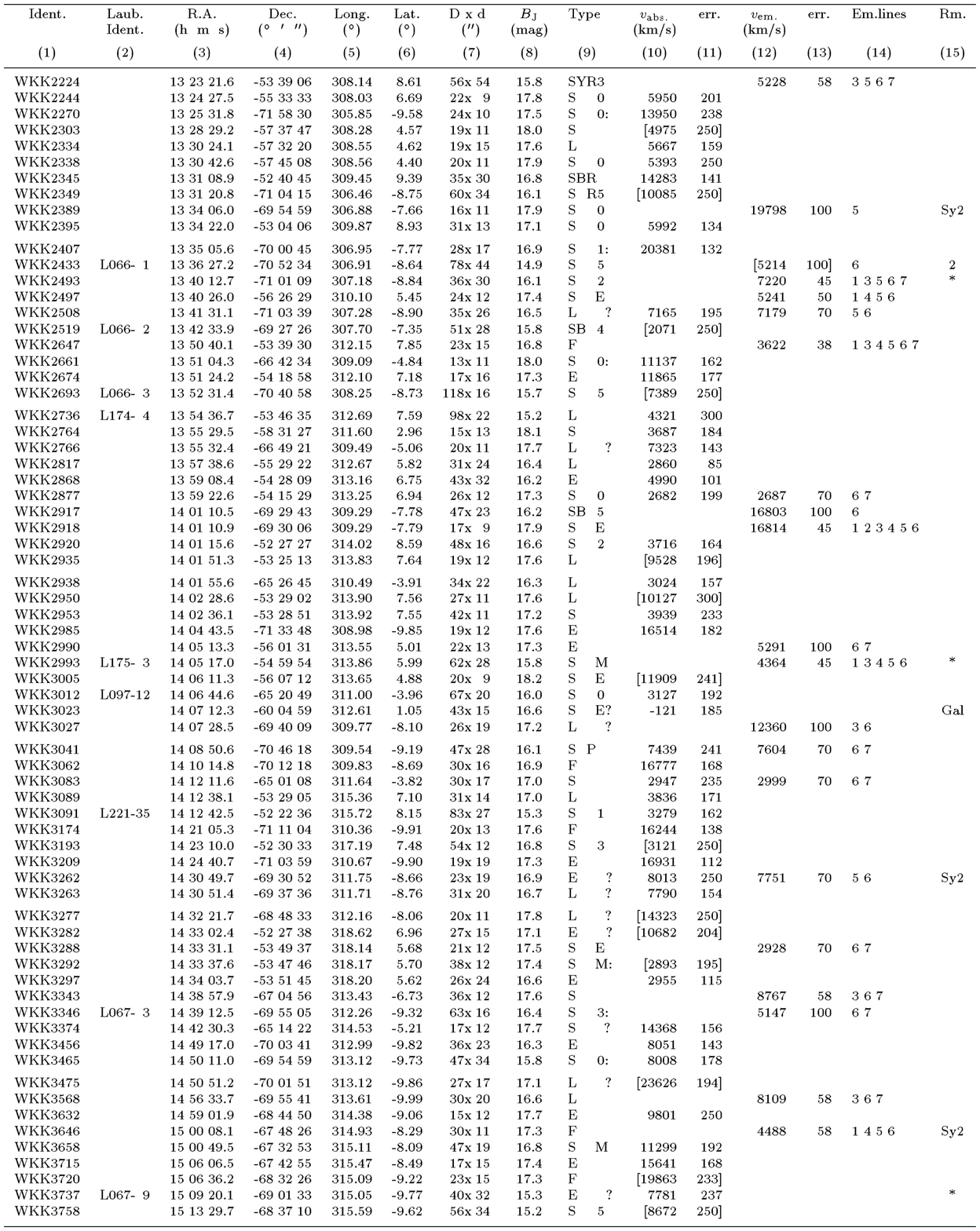}\hfil
\end{table*}

\begin{table*}
\caption{Galaxies with redshifts obtained at S.A.A.O. in the Hydra/Antlia Region, addendum Paper I.}
\hfil\epsfxsize 17.5cm \epsfbox[31 272 600 580]{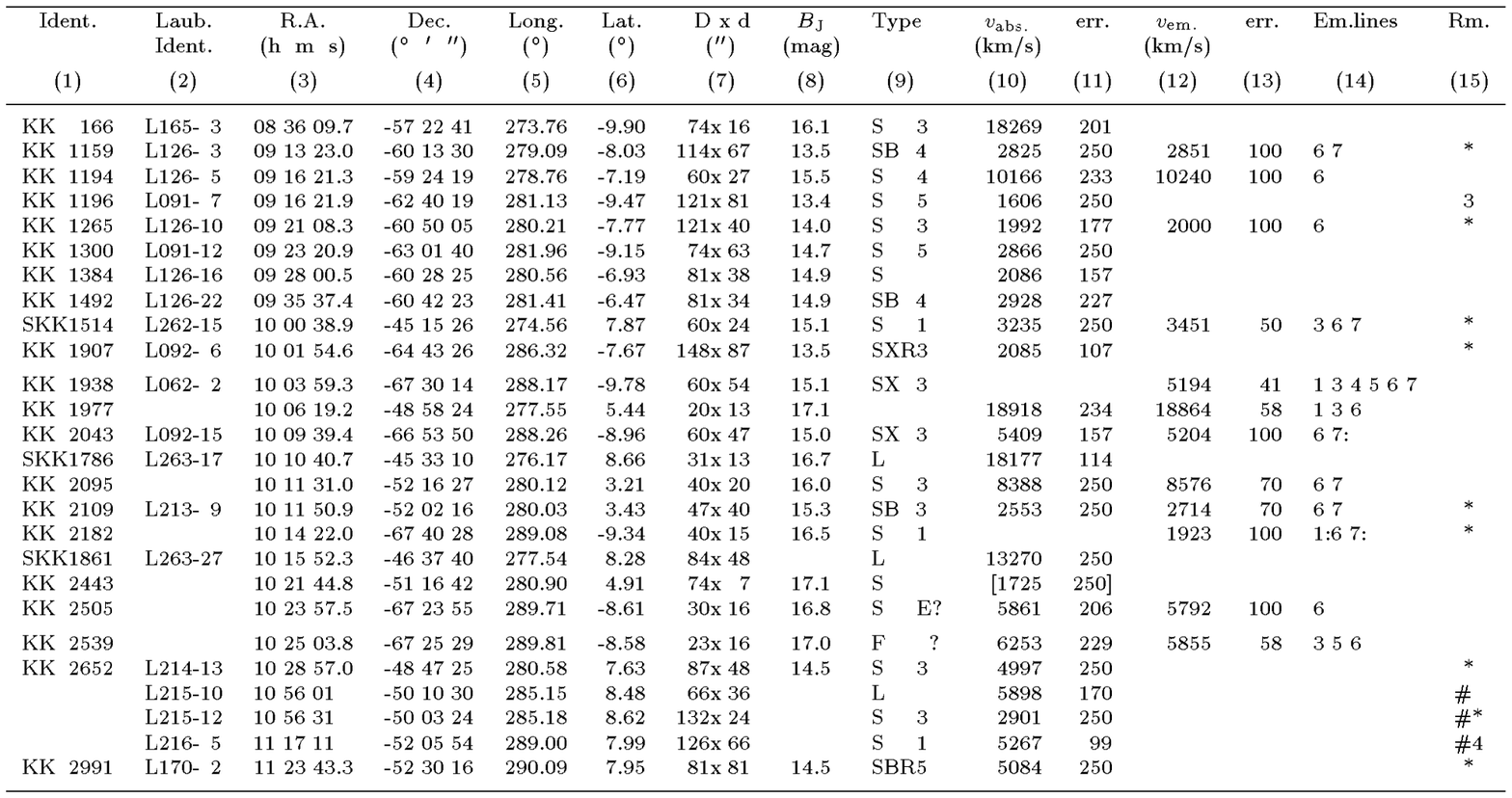}\hfil
\end{table*}

The entries in Table~1 are as follows:
\begin{description}

\item [Column 1 and 2:] Identification of the galaxy as given in WKK97
and Lauberts identification (Lauberts, 1982). 

\item [Column 3 and 4:] Right Ascension and Declination (1950.0). The
positions were measured with  the Optronics machine at the 
ESO in Garching and have an accuracy of about 1 arcsec.

\item [Column 5 and 6:] Galactic longitude $\ell$
and latitude $b$.

\item [Column 7:] Large and small diameters (in arcsec). The diameters
are measured approximately to the isophote of 24.5 mag arcsec$^{-2}$ and
have a scatter of $\sigma \approx 4\arcsec$.

\item [Column 8:] Apparent magnitude B${\rm_J}$. The magnitudes are estimates
from the film copies of the SRC IIIaJ Survey based on the above given 
diameters and an estimate of the average surface brightness of the galaxy. 
The magnitudes and diameters in the Crux region 
are estimated by PAW but show no significant deviation from the magnitude and
diameter estimates in the Hydra--Antlia region as determined by RCKK 
(Paper I and Table 2). A more detailed discussion on this 
subject will follow in WKK97.

\item [Column 9:] Morphological type. The morphological types are coded 
similarly to the precepts of the Second Reference Catalogue 
(de Vaucouleurs \etal 1976). Due to 
the varying foreground extinction a homogenous and detailed type 
classification could not always be accomplished and some
codes were added: 
In the first column F for E/S0 was added to the normal designations
of E, L, S and I. In the fourth column the subtypes E, M and L are 
introduced next to the general subtypes 0 to 9. They stand for
early spiral (S0/a-Sab), middle spiral (Sb-Sd) and late spiral or 
irregular (Sdm-Im). The cruder subtypes are a direct indication of the 
fewer details visible in the obscured galaxy image. The 
questionmark at the end marks uncertainty of the main type, the colon
uncertainty in the subtype.

\item [Column 10 and 11:] Heliocentric velocity (cz)
and error as derived from the absorption features. The errors may
appear large as they are estimated external errors, and not internal errors
(see Paper I). The square brackets indicate a tentative redshift.

\item [Column 12 and 13:] Heliocentric velocity and error measured
from the emission lines (identified in column 14) when present.
The square brackets indicate a tentative redshift.

\item [Column 14:] Identified emission lines: \hfill \phantom{a}

\begin{tabular}{ccccccc}
 1 & 2 & 3 & 4 & 5 & 6 & 7 \\
 $[$OII] & H$\gamma$ & H$\beta$ & [OIII] & [OIII] & H$\alpha$ & [NII] \\
 3727 & 4340 & 4861 & 4959 & 5007 & 6563 & 6584 \\
\end{tabular}
A colon indicates the identification is tentative.
\end{description}

\begin{table*}
\caption{Galaxies observed at the SAAO without reliable redshift.}
\hfil\epsfxsize 18cm \epsfbox[31 317 600 536]{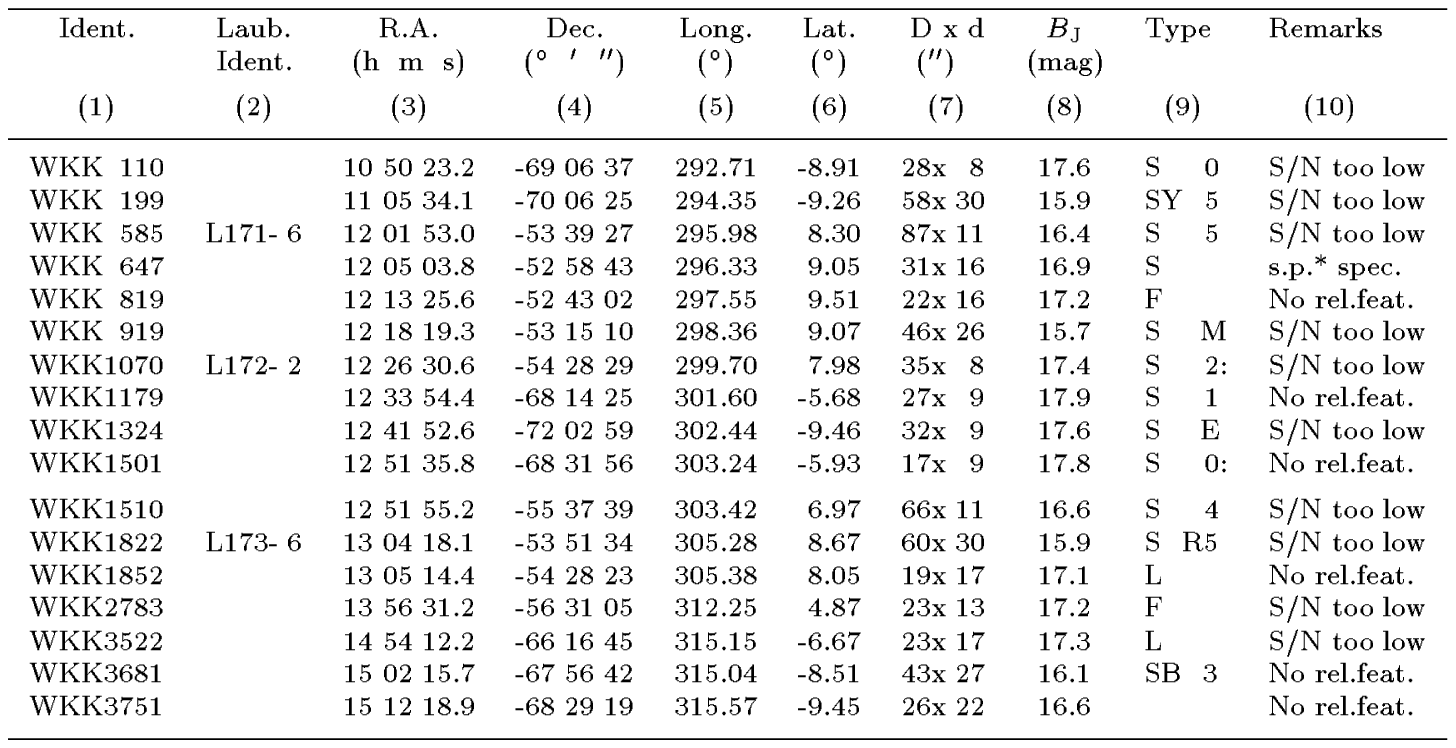}\hfil
\end{table*}


\begin{table*}
\caption{Galaxies in the Crux region with radial velocities from the literature (NED) in the Crux Region.}
\hfil\epsfxsize 18cm \epsfbox[31 131 600 702]{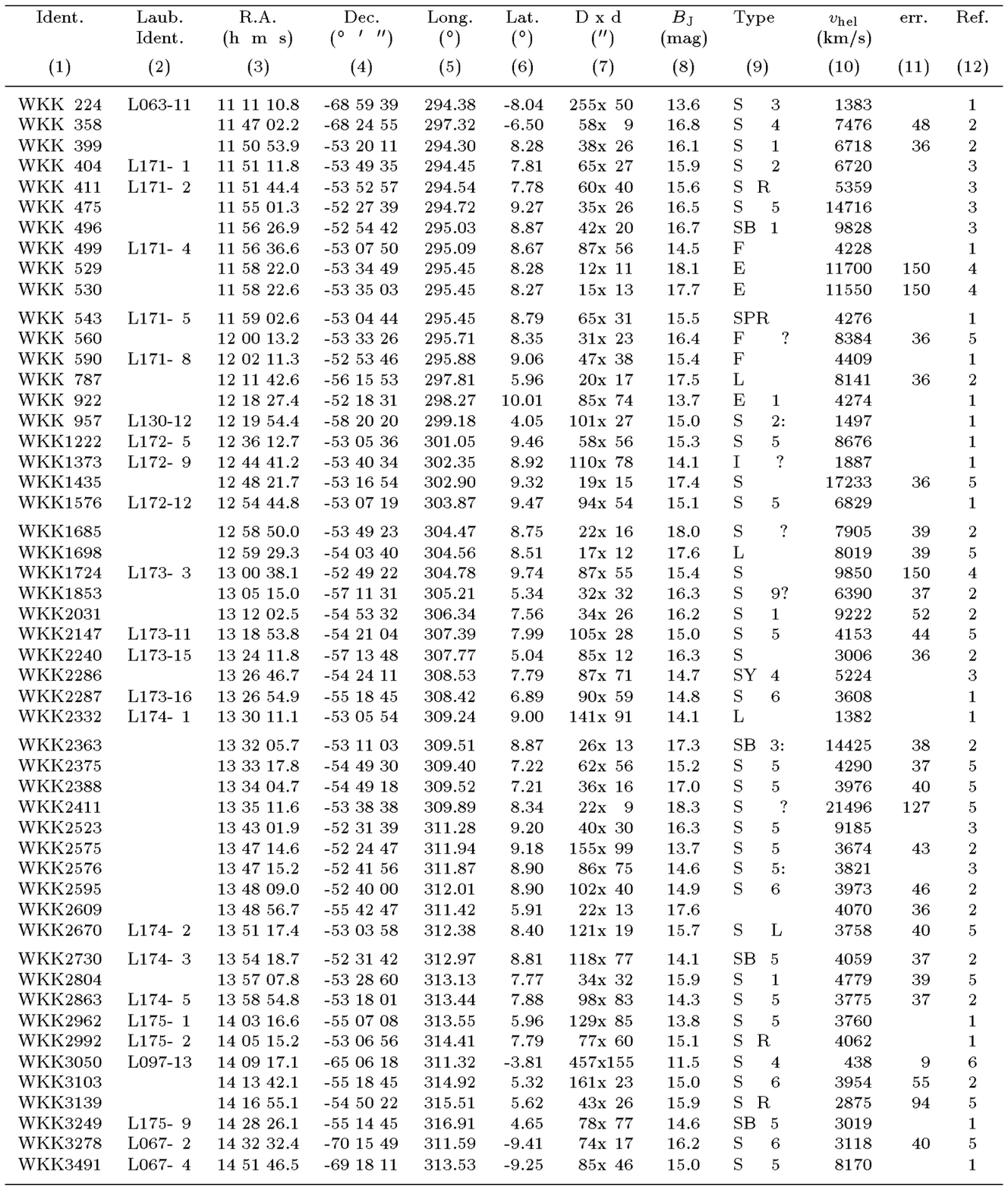}\hfil
\end{table*}

\begin{description}
\item [Column 15:] Code for additional remarks:\\
1 -- \hspace{0.15cm} WKK 150: The redshift measured at the SAAO for this galaxy is 
in disagreement with the value quoted in the literature ($v = 8948 \pm 40$ {\kms}, Fisher et 
al. 1995). It might be due to an identification error; a neighbouring galaxy (WKK 158) was
found to be in good agreement with the value quoted in the literature.\\
2 -- \hspace{0.15cm} The redshifts for these galaxies are tentative only, but confirmed by 
independent redshifts from the literature.\\
Sy2 -- \hspace{0.15cm} These four galaxies have been classified as Seyfert 2.\\
PN -- \hspace{0.15cm} WKK P8: This object was classified as a possible PN, but was 
not listed in PNe list of Acker 1992. It has now been confirmed as a PN.\\
Gal -- \hspace{0.15cm} Identification as galaxy questionable, spectra indicates a galactic 
origin.\\
$*$ -- \hspace{0.15cm} Redshifts are also available in the literature.\\
\end{description}

Table~2 represents an addendum to our Paper I, giving further redshift
measurements for galaxies in the Hydra--Antlia region. The format is the same 
as for Table~1 and the entries in Table 2 are as in Table 1 with the exception of:

\begin{description}

\item [Column 1:] Identification of the galaxy as given in Kraan-Korteweg (1997) for 
galaxies with the prefix KK, or from Salem \& Kraan-Korteweg (1997) for galaxies with 
the prefix SKK. 

\item [Column 8:] Apparent magnitude B${\rm_J}$ as determined by RCKK. These values are
derived similar to the ones quoted in Paper I.

\item [Column 15:] Code for additional remarks:\\
3 -- \hspace{0.15cm} KK1196: The redshift measured at the SAAO for this galaxy is
in slight disagreement with the value quoted in the literature  ($v = 2411$ {\kms}, 
Huchtmeier \& Richter 1989).\\
4 -- \hspace{0.15cm} L216- 5:  The redshift measured at the SAAO for this galaxy is
in slight disagreement with the value quoted in the literature  ($v = 6350$ {\kms}, 
Visvanathan \& v.d.~Bergh 1992).\\
\# -- \hspace{0.15cm} These galaxies are not in our search area, but do lie in the
``Zone of Avoidance''. The diameters and type are from Lauberts 1982.\\
\end{description}

\noindent{\bf Special cases:}

In the course of this survey we have given special attention to a number of objects that,
based on their optical appearance, might be Planetary Nebulae (PNe). In the Crux region 
there a four such objects and these observations are described elsewhere
(Kraan-Korteweg et al. 1996b): 
Two of the four PNe candidates were too faint to be detected, one was confirmed
as being a PN (PNG 299.5+02.5) and one object was classified as a new PN 
(PNG 298.3+06.7).

Table~3 lists galaxies for which the spectra has too poor a signal-to-noise
ratio for a redshift to be extracted. For one galaxy (WKK 647) the spectrum was
dominated by a superimposed foreground star and no reliable redshift could be 
extracted.

Table~4 lists the galaxies with their optical properties as determined
in our Crux survey, for which redshifts have already been published in 
the literature. Although now reobserved by us, these galaxies clearly
would have been included in
our observations, since they meet our selection criteria.
Columns 1-9 are the same as in Table~1. 
Column 10 and 11 list the heliocentric
velocities and errors (if given). The velocity
in column 10 has been adopted from the source identified in
column 12, where the number corresponds to:

\begin{enumerate}
\item{Dressler 1991, The supergalactic plane redshift survey.}
\item{Strauss et al. 1992, A redshift survey of IRAS galaxies.}
\item{Visvanathan \& v.d.~Bergh 1992, Luminous spiral galaxies in the direction of the 
Great Attractor.}
\item{Fairall 1983, Spectroscopic survey of southern compact and bright-nucleus
galaxies.}
\item{Fisher et al. 1995, The IRAS 1.2 Jy Survey.}
\item{Fairall 1996, The Southern Redshift Catalogue.}
\end{enumerate}

\subsection{Comparison to other measurements}

Although the galaxies we observed were initially selected on the basis of having 
no published redshifts, more recent publications have provided redshifts 
from other investigators for 20 of the galaxies in Table 1 and Table 2.  This overlap, 
however, allows a suitable comparison between our sample and others.  
We find
\begin{displaymath}
<v_{SAAO} - v_{pub}> = +7 \pm 94 {\rm \; km \, s^{-1}}
\end{displaymath}  
 
which shows no significant systematic error, and agrees well 
within our average standard deviation.
 
Similarly, we have allowed a small overlap between the SAAO 
galaxies and our complementary programmes using Mefos and Parkes 
radio observations, for which we find

\begin{displaymath}
<v_{SAAO} - v_{MEFOS}> = -49 \pm 118 {\rm \; km \, s^{-1}}
\end{displaymath}
\begin{displaymath}
<v_{SAAO} - v_{PKS}> = +49 \pm 135 {\rm \; km \, s^{-1}}.
\end{displaymath}
 
The comparison with the MEFOS and Parkes data is based on 5 and 12 galaxies 
respectively. Again, the agreement is acceptable with our standard deviation; 
the latter in any case probably reflects low surface-brightness 
galaxies, for which our errors are at their largest values.

\section{Coverage and Completeness}
 
We have obtained reliable redshifts for 173 galaxies, with tentative
redshifts for a further 36 objects, as set out in Table 1. The tentative 
redshifts are not used in any of the subsequent plots or analyses. 

\begin{figure}
\hfil\epsfxsize 8.8cm \epsfbox{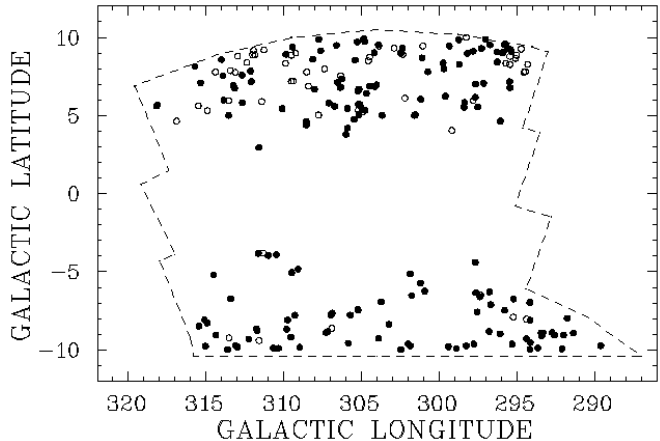}\hfil
\caption{Distribution of the galaxies with radial velocities. 
The Crux survey region is outlined.
Solid circles indicate the positions of galaxies, in the Crux region, 
observed in the present work, and for which redshifts have been obtained.  
Open circles show the positions of galaxies for which redshifts are 
available from the literature.}
\label{f2}
\end{figure}

\begin{figure}
\hfil\epsfxsize 8.8cm \epsfbox{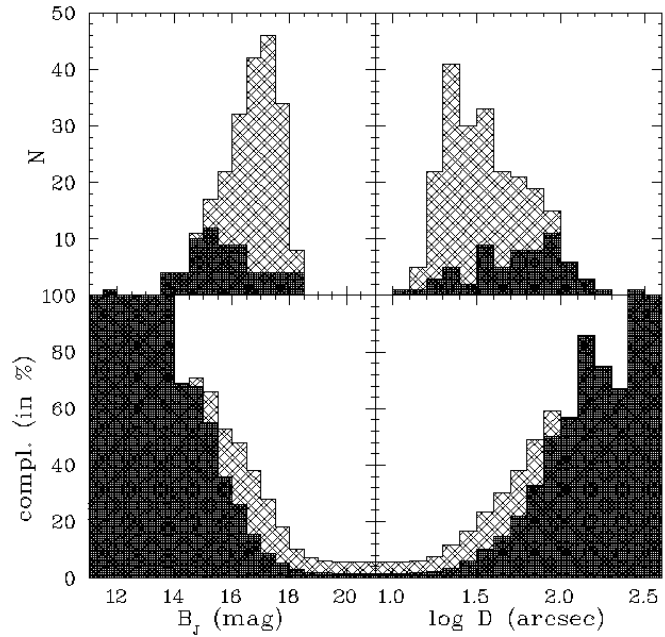}\hfil
\caption{Magnitude and major-axis diameter distribution of galaxies with 
radial velocities in the Crux search area. The lighter hatched areas mark 
galaxies observed at the SAAO, the darker hatched by others. The bottom panel
shows the completeness in percent of the total number of galaxies found in the
Crux Region (cf.~fig.~1).}
\label{f3}   
\end{figure}

Figure 2 shows the sky coverage of the galaxies with reliable redshifts, depicted
by solid circles. The coverage can be seen to be reasonably uniform, on
both the north and south sides of the galactic obscuration. By contrast, 
the data from Table 4 -- galaxies with previously known redshifts as depicted
by the open circles --
is strongly concentrated north of the plane. This is due to the 
redshift surveys in that area that focussed on the Great Attractor (esp. 
Dressler 1991, Visvanathan \& van den Bergh 1992).

Figure 3 presents a comparison of the magnitudes and major-axis 
distribution of the observed galaxies; it shows general similarity 
to the corresponding plot in Paper I. 
The sharp magnitude cut-off indicates the limiting performance of 
the 1.9m telescope for galaxies fainter than (B$_{\rm{J}} \la 18\mag0$).

These observations trace the bright end of the magnitude distribution of the
3760 galaxies in the Crux region very well. We are 70\% complete for galaxies
brighter than (B$_{\rm{J}} \la 14\mag5$) and even 54\% complete for galaxies
brighter than (B$_{\rm{J}} \la 16\mag0$). These numbers are comparable to
the completeness figures in the Hydra--Antlia region. 
As before, a number of bright galaxies are missed by this survey. They are
extended low surface brightness spiral galaxies. In the meantime, they 
have been observed with the Parkes 64--m telescope and will be reported on
elsewhere.

\section {Identification of Large-Scale Structures}

\subsection{Velocity distribution}

The histogram of redshifts shown in Figure 4 reveals a broad 
concentration of galaxies around $3500 \le v \le 8500$ {\kms}, with a probable 
dip at $4500$ {\kms}.  This is quite different to the single peak 
centred at $2750$ \kms for the neighbouring Hydra--Antlia region 
to the west (Paper I), and the dominating peak at $4882$ \kms of 
the Norma cluster (KK96) to the east.  It suggests that quite 
different large-scale structures may be present in this Crux region, 
which we shall examine in greater detail below.    

\begin{figure}
\hfil\epsfxsize 8.8cm \epsfbox{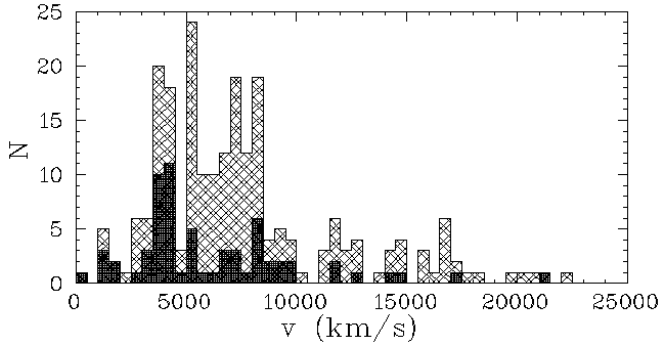}\hfil
\caption{Velocity histogram of the galaxies in the search area
in the Crux extension in the ZOA. Lighter hatched areas are velocities
measured by us; darker hatched are previous observations by others.}
\label{f4}
\end{figure}

\subsection{Sky projection}

In regard to the interpretation of data in three-dimensional redshift 
space, particularly in the absence of magnitude controls, the reader 
is directed to the introductory discussion in Section 3.2 of Paper I.
In interpreting these plots, note that the Hydra--Antlia ZOA region
is covered observationally as deep as the current Crux region, whereas
other low--latitude regions only have very sparse redshift information

\begin{figure*}
\hfil\epsfysize 21cm \epsfbox{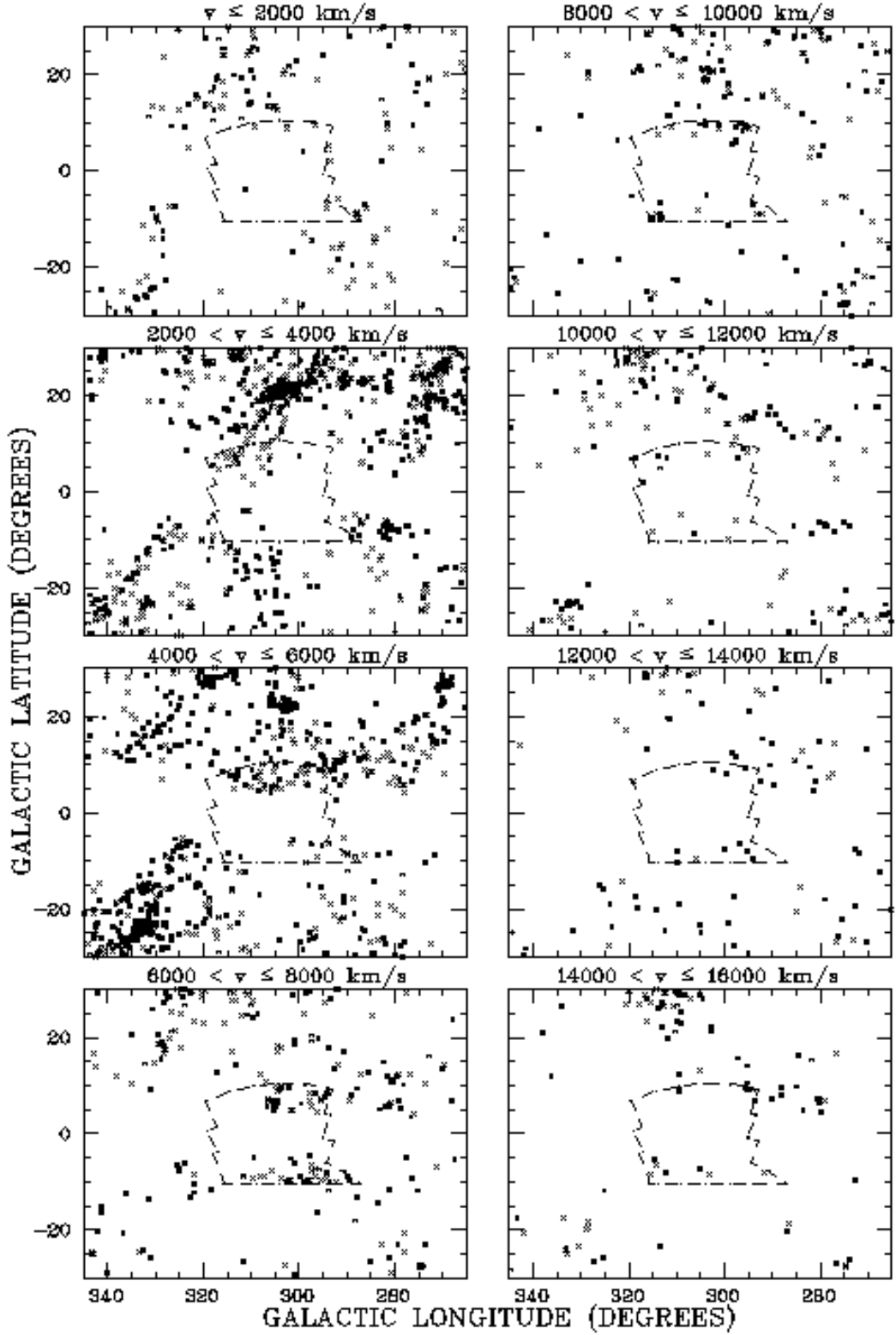}\hfil
\caption{Sky projections in galactic coordinates for redshift intervals
of $\Delta v=2000$ {\kms}. Within the panels the redshifts are subdivided
into intervals of $\Delta v=1000$ {\kms}: filled squares mark the nearer
redshift interval (\eg $v\le1000$ {\kms} in the top-left panel),
crosses the more distant interval ($1000<v\le 2000$ \kms in same panel).
The skyplots increase in velocity-distance from the top-left panel to 
the bottom-right panel as marked above each panel.
The area of our investigation is outlined.}
\label{f5}
\end{figure*}

Figure 5 shows the surveyed region and its surrounding volume sliced 
in redshift intervals of $\Delta v = 2000$ {\kms}.  As already indicated, 
most of the data points, within the outlined Crux region, are from the 
new observations. Conspicuous features occur in the second and third slices
 - matching the excess already noted in the histogram of Figure 4.

In the second slice, the new data reinforce the presence of a narrow filamentary 
structure running all the way from $\ell = 340\deg$, $b = -25\deg$  
to the Centaurus cluster at $\ell = 303\deg$, $b = 20\deg$.
This is believed to be part of a Great Wall-like structure seen edge-on
-- the Centaurus Wall.  
The new observations fill in a portion of 
this feature previously hidden by the Milky Way - as might be expected of a 
continuous massive structure.  
The break, due to the obscuration close to the galactic plane 
is now much narrower.  The new data in the top-left corner of the Crux
region is concentrated in the $3500-4500$ \kms
range - slightly closer than most other condensations within the wall.
 There are also indications of weaker filaments: One crosses north-south 
at $\ell = 305\deg$ while another passes just outside the south-east corner
of the surveyed area.  

However, the most important structure revealed by the present survey 
occurs in the third slice down.  This is a concentration of galaxies 
centred at $5000$ \kms in the upper (northern) segment of the surveyed 
volume.  Together with the neighbouring galaxies outside the survey 
volume, it suggests a large-scale structure running more or less 
horizontally across the diagram. We have earlier labelled this structure 
as the ``Norma supercluster'' (Woudt et al. 1997). Traces can also be seen in the following 
slice, so the feature is also probably wall-like seen 
roughly side on - i.e. its width (or depth in Fig 5) being some $3000$ \kms 
and its thickness several hundred {\kms} unless much is still hidden by 
the dense obscuration.  We shall see however (in Fig 6) that this 
feature is more complex.  For the moment note the concentration at 
$\ell = 305\deg$ in the fourth slice - coincident with that seen earlier in 
Figure 1.  Bearing in mind its greater distance, compared to the 
Centaurus Wall mentioned above, this new structure must be similarly 
massive.
The Norma cluster is situated where these two massive structures -- the 
Centaurus Wall and the Norma supercluster -- intersect.

Features at greater redshifts are understandably narrower in angular 
dimensions, and more difficult to discern: The fourth slice shows a 
feature running along the southern boundary of the Crux region.  
Beyond that, the data is generally too sparse, except for a weak concentration 
at $14000 - 16000$ \kms in the south-east (bottom left), that corresponds 
to the relatively distant overdensity noted earlier.

\begin{figure}  
\hfil\epsfxsize 8.8cm \epsfbox{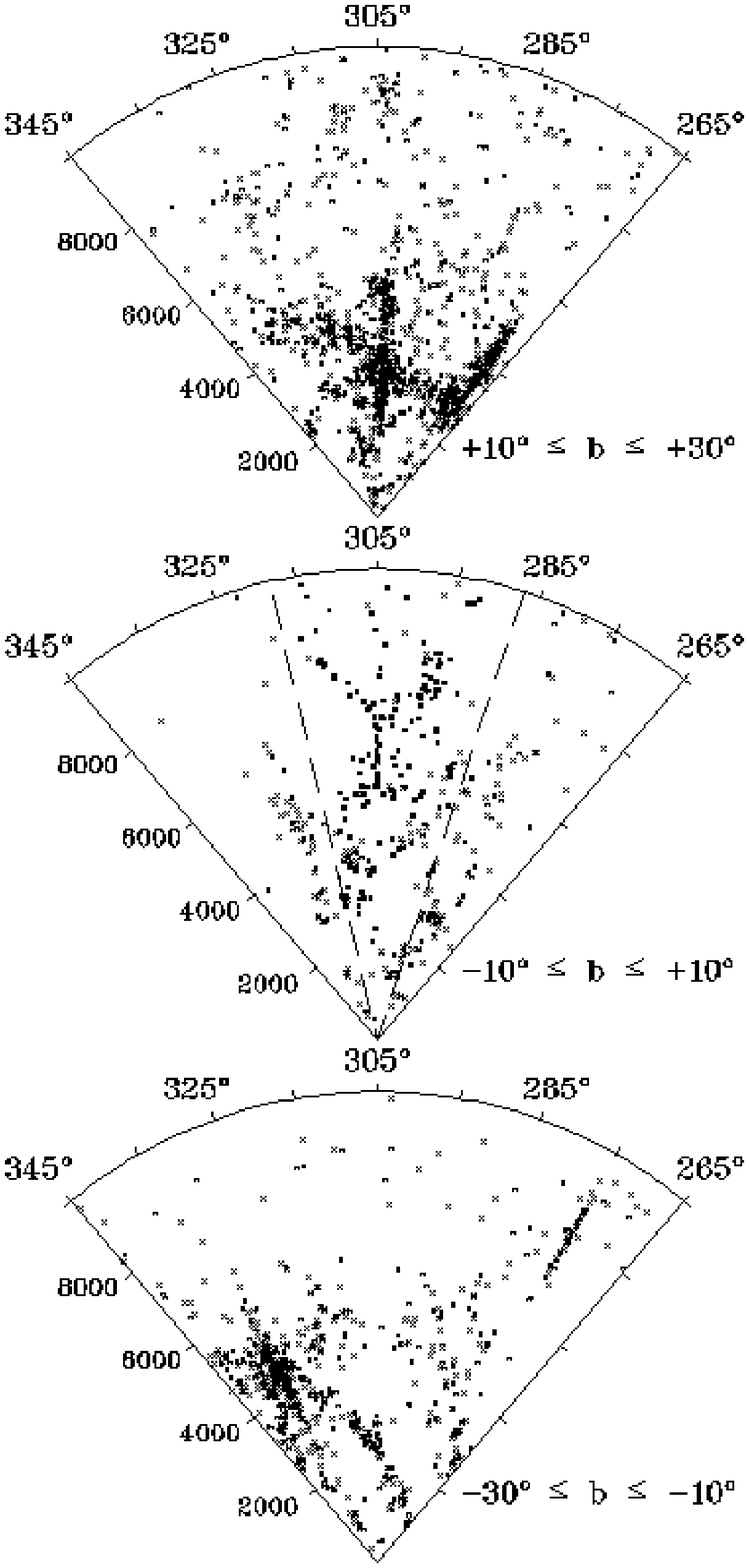}\hfil
\caption{Redshift slices out to $v < 10000$ \kms 
for the longitude range $265\deg < \ell <
345\deg$. The top panels display the structures above
the GP ($+10\deg < b \le +30\deg$) the middle panel in the GP 
($-10\deg \le b \le +10\deg$) and the bottom panel the structures below
the GP ($-30\deg < b \le -10\deg$). The dashed lines in the middle panel
delimits the survey area. Filled squares are measurements from
the SAAO, crosses from the literature.}
\label{f6}
\end{figure}

\subsection{Pie diagrams}

Our final set of plots, Figure 6, shows slices in galactic latitude; 
the uppermost including the Centaurus and Hydra clusters, the lowest 
the Pavo cluster.
 
The product of our survey appears as the middle slice, which would 
otherwise be largely blank.  Instead, the slice reveals a cellular 
structure - familiar from plots elsewhere in the sky (eg Fairall et 
al. 1990) - but never before discerned so clearly at so low a galactic 
latitude. Presumably, the appearance is somewhat assisted by a 
favourable angle of cut.  The cell sizes, and the voids so contained, 
are $1000-2000$ \kms in diameter.
 
Of particular interest is a radial feature (from 5300 to 7000 \kms), 
almost dead centre, that corresponds to the 
concentration at galactic longitude $\ell = 305\deg$.  The feature
could be the ``finger of God'' of a small cluster except it is not
discernable as such in a central longitude slice (not reproduced here).
 
\section {Summary.}

The main finding of this survey has been an overdensity of galaxies 
in the $3000 - 8000$ \kms range in redshift.  Rather than a single 
feature, this overdensity represents a set of distinct cellular 
structures similar to those seen at higher galactic latitudes
and for which Figure 6 (middle plot) provides a mapping. The survey
also provides more detailed mapping of the ``Centaurus Wall'' where it 
crosses the Milky Way.

\acknowledgements
{The authors would like to thank the night assistants Francois van Wyk
and Fred Marang as well as the staff at the SAAO for their hospitality. 
APF and PAW are supported by the South African FRD. The research by RCKK is
being supported with an EC-grant. Financial support was provided by 
CNRS through the Cosmology GDR program.
This research has made use of the NASA/IPAC
Extragalactic Database (NED), which is operated by the Jet Propulsion 
Laboratory, Caltech, under contract with the National Aeronautics and Space
Administration.
}

\end{document}